\documentclass[pra,showpacs,amsmath,amssymb,amsfonts,lengthcheck,longbibliography,superscriptaddress]{revtex4-2}

\usepackage{upgreek} 

\usepackage{changes}
\setdeletedmarkup{\color{blue}\sout{#1}}
\usepackage{cancel}

\usepackage[utf8]{inputenc}
\usepackage[T1]{fontenc}
\usepackage{qcircuit}

\usepackage{mathrsfs} 
\usepackage[normalem]{ulem}
\usepackage{graphicx, color, graphpap}
\usepackage{enumitem}
\usepackage{amssymb}
\usepackage{amsthm}
\usepackage{multirow}
\usepackage[colorlinks=true,citecolor=blue,linkcolor=blue,urlcolor=blue]{hyperref}
\usepackage[T1]{fontenc}
\usepackage{verbatim}
\usepackage{mathtools}
\usepackage{titlesec}

\long\def\ca#1\cb{} 

\newcommand{\abs}[2][]{#1| #2 #1|}

\newcommand{\bramatket}[3]{\langle #1 \hspace{1pt} | #2 | \hspace{1pt} #3 \rangle}

\newcommand{\avg}[1]{\langle #1\rangle }
\newcommand{\ket}[1]{|#1\rangle}               
\newcommand{\bra}[1]{\langle #1|}              
\newcommand{\dya}[1]{\ket{#1}\!\bra{#1}}

\newcommand{\tr}[1]{\mathrm{tr}\left\{#1\right\}}

\newcommand{\ave}[1]{\langle #1\rangle}               

\renewcommand{\Re}{\text{Re}}
\renewcommand{\Im}{\text{Im}}

\newcommand{\ad}{^\dagger}

\newcommand*{\id}{\openone}

\newcommand{\eq}{\text{eq}}
{}
{}

\begin{document}
\title{Detailed fluctuation theorem {from the} one-time measurement scheme}

\author{Kenji Maeda}
\affiliation{Department of Physics, University of Massachusetts, Boston, Massachusetts 02125, USA}

\author{Tharon Holdsworth}
\affiliation{Department of Physics, University of Massachusetts, Boston, Massachusetts 02125, USA}

\author{Sebastian Deffner}
\affiliation{Department of Physics, University of Maryland, Baltimore County, Baltimore, Maryland 21250, USA}

\author{Akira Sone}
\email{akira.sone@umb.edu}
\affiliation{Department of Physics, University of Massachusetts, Boston, Massachusetts 02125, USA}

\begin{abstract} 
We study the quantum fluctuation theorem in the one-time measurement (OTM) scheme, where the work distribution of the backward process has been lacking and which is considered to be more informative than the two-time measurement (TTM) scheme. We find that the OTM scheme is the quantum nondemolition TTM scheme, in which the final state is a pointer state of the second measurement whose Hamiltonian is conditioned on the first measurement outcome. Then, by clarifying the backward work distribution in the OTM scheme, we derive the detailed fluctuation theorem in the OTM scheme for the characteristic functions of the forward and backward work distributions, which captures the detailed information about the irreversibility and can be applied to quantum thermometry. We also verified our conceptual findings with the IBM quantum computer. Our result clarifies that the laws of thermodynamics at the nanoscale are dependent on the choice of the measurement and may  provide experimentalists with a concrete strategy to explore laws of thermodynamics at the nanoscale by protecting quantum coherence and correlations. 
   
\end{abstract}

\maketitle

One of the most significant conceptual factors distinguishing quantum physics and classical physics is \textit{measurement}~\cite{Nielsen}. In quantum mechanics, measurements {typically} destroy quantum coherences and correlations that {could} be utilized as the resources for many quantum engineering tasks, such as quantum computing and quantum metrology. Compared to classical systems, one has many degrees of freedom in choosing the basis of the measurement based on their task on the quantum system. Particularly, the {eigenbasis} of the observable of the measurement apparatus {is comprised of the so-called} pointer states~\cite{zurek1981pointer, schlosshauer2007decoherence, brasil2015understanding, touil2022eavesdropping,zurek2003decoherence}, {which are immune to} decoherence due to the corresponding measurement.

Quantum thermodynamics~\cite{DeffnerBook19,Anders16,Binder19} is a rapidly growing field {exploring} the laws of thermodynamics from the perspective of quantum information science. Fluctuation theorems~\cite{Jarzynski97,crooks1999entropy} in both quantum and classical systems are regarded as one of the most significant laws to date~\cite{Ortiz2011} because many significant thermodynamic principles can be derived, such as the second law of thermodynamics~\cite{jarzynski2011equalities} and response theory~\cite{andrieux2008quantum,andrieux2009fluctuation}. The standard approach toward quantum fluctuation theorem is the so-called two-time measurement (TTM) scheme~\cite{Tasaki00,Kurchan01,Talkner2007,Jarzynski04,Morikuni17,Rastegin13,Jarzynski15,Zhu16,Pan19,Kafri12,Goold15,Rastegin14,Goold14,Acin2017NoGo,Gardas2018,kiely2023entropy}. 

{The} TTM scheme is constructed by two energy projection measurements at {the beginning and the end} of {a} quantum process. In the standard setup of the time-varying Hamiltonian system, the initial state is prepared in the Gibbs state defined by its initial Hamiltonian $H_0$. Then, one performs an energy measurement on the initial state with $H_0$, which projects the system onto one of the eigenstates $\ket{E_i}$ of $H_0$ based on the initial measurement outcome $E_i$. Then, one evolves the system under the unitary operator $U$ during time $\tau$ and measures the evolved state $U\ket{E_i}$ with the final Hamiltonian $H_{\tau}$. {Finally}, the system will be projected again onto an eigenstate $\ket{E_j'}$ of $H_{\tau}$ based on the final measurement outcome $E_j'$.  

The work performed on the system in a single {realization} is defined by the difference between the final and initial measurement outcome, $W_{i\to j}\equiv E_j'-E_i$ , which recovers the standard fluctuation theorem, also known as the TTM fluctuation theorem {resembling the classical Jarzynksi equality} \cite{Jarzynski97}. Therefore, the TTM scheme can be regarded as a \textit{semiclassical} approach, which has been experimentally implemented in various systems~\cite{Huber2008,Smith2018,An15,Tiago14,hernandez2020experimental,hernandez2021experimental,Collin05,Sagawa15,Sagawa10b,Zhenxing18,hahn2022verification}, {including a demonstration on the DWave machine \cite{Gardas2018}.} However, the second projection measurement usually destroys the quantum coherence and correlations generated through the dynamics, which means that the TTM cannot fully capture the peculiar features of the quantum systems when one analyzes its thermodynamic behaviors~\cite{Acin2017NoGo}.

{To address the thermodynamic contribution of quantum correlations,} Ref.~\cite{Deffner16} proposed the so-called one-time measurement (OTM) scheme. In this scheme, the second measurement is considered to be avoided, and the work is determined by the energy difference conditioned on the initial energy measurement outcome. {Within this paradigm, the corresponding} Jarzynski equality includes the additional information contribution {stemming} from the quantum relative entropy of the conditional thermal state~\cite{Sone21b,sone2023conditional} with respect to the Gibbs state defined by the final Hamiltonian. This additional term provides a tighter maximum work relation and captures the quantum coherence or correlations generated through the dynamics in the formalism. Therefore, the OTM scheme {can be regarded as} more informative than the TTM scheme. {This has been elucidated} in various contexts, including quantum thermometry~\cite{sone2023conditional}, work as an external quantum observable~\cite{Beyer2020}, distinguishability of heat and work in an open quantum system~\cite{Sone20a}, heat exchange~\cite{sone2022heat}, classical correspondence of {the} OTM scheme~\cite{Sone21b}, quantum ergotropy ~\cite{sone21a}, and information production~\cite{sone2023jarzynski}. However, the backward process in the OTM scheme has not been considered yet, which has made the detailed quantum fluctuation theorem of the OTM scheme {elusive}.

In the present Letter, we first prove that the OTM scheme is the  {quantum nondemolition (QND)} TTM scheme, where the pointer states of the second measurement (conditional Hamiltonian) are the evolved states conditioned on the initial measurement outcome. From this, we construct the backward work distribution and derive the detailed quantum fluctuation theorem of the OTM scheme, which we call \textit{OTM fluctuation theorem}. Then, we propose a quantum circuit to compute the symmetric relation of the characteristic functions of the forward and backward work distributions.  {We explore the physical meaning of the OTM fluctuation theorem by associating it to the concept of irreversibility and demonstrate the potential application of the derived formalism to state preparation for low-temperature quantum thermometry.}
Finally, we verify the derived detailed fluctuation theorem with IBM quantum computer to demonstrate the experimental implementability of the OTM scheme. These results {emphasize} that the laws of {quantum} thermodynamics are strictly determined by the choice of measurements by the observers.

\subparagraph{OTM detailed fluctuation theorem}
Our first result is the derivation of the OTM fluctuation theorem. We consider a finite-dimensional closed quantum system described by a $d$-dimensional Hilbert space. Let the initial state be a Gibbs state $\rho_0^{\eq}\equiv \exp(-\beta H_0)/Z_0$, where $H_0$ is the initial Hamiltonian and $Z_0\equiv\tr{\exp(-\beta H_0)}$ is the partition function. In {a} closed quantum system, the time evolution is described by a unitary operator $U$.  In the OTM scheme, the work {for a single realization of the protocol} is defined as 
\begin{equation}
    \widetilde{W}_i \equiv \bramatket{E_i}{U\ad H_{\tau} U}{E_i}-E_i,
\label{eq:ConditionalWork}
\end{equation}
which {also has been} called \textit{conditional work}~\cite{Sone21b}. This is the energy difference between the final energy conditioned on the initial measurement outcome and itself. Then, the forward conditional work distribution is {simply} given by~\cite{Deffner16}
\begin{equation}
    \widetilde{P}_f(W) = \sum_{i=1}^{d}\frac{e^{-\beta E_i}}{Z_0}\delta\left(W-\widetilde{W}_i\right),
\label{eq:ForwardW}
\end{equation}
which {is consistent with} the exact average work 
\begin{equation}
\ave{W}=\int W\widetilde{P}_f(W)dW=\tr{\left(U\ad H_{\tau}U-H_0\right)\rho_0^{\eq}}
\end{equation}
and {yields} the {generalized} Jarzynski equality \cite{Deffner16}
\begin{equation}
\label{eq:gen_jar}
\ave{e^{-\beta W}}_{\widetilde{P}}=\frac{\widetilde{Z}_{\tau}}{Z_0}=e^{-\beta\Delta F}e^{-S(\widetilde{\rho}_{\tau}||\rho_{\tau}^{\eq})}.
\end{equation}
{In Eq.~\eqref{eq:gen_jar}, the conditional partition function, $\widetilde{Z}_{\tau} \equiv \sum_{i=1}^{d} \exp{\left(-\beta\bramatket{E_i}{U\ad H_{\tau} U}{E_i}\right)}$,} is the normalization factor used to construct the conditional thermal state,
\begin{equation}
    \widetilde{\rho}_{\tau}\equiv \sum_{i=1}^{d} \frac{e^{-\beta\bramatket{E_i}{U\ad H_{\tau} U}{E_i}}}{\widetilde{Z}_{\tau}} U\dya{E_i}U\ad.
\end{equation}
{Finally}, $S(\widetilde{\rho}_{\tau}||\rho_{\tau}^{\eq}) =  \tr{\widetilde{\rho}_{\tau} \ln \widetilde{\rho}_{\tau}}-\tr{\widetilde{\rho}_{\tau} \ln \rho_{\tau}^{\eq}}$ is the quantum relative entropy of the conditional thermal state with respect to the Gibbs state $\rho_{\tau}^{\eq}\equiv\exp(-\beta H_{\tau})/Z_{\tau}$ of the final Hamiltonian $H_{\tau}$.

By comparing with the TTM scheme, we demonstrate that the OTM scheme is equivalent to the TTM scheme with a carefully chosen final Hamiltonian (conditional Hamiltonian) based on the information about the initial measurement outcome and the dynamics of the system. To see this point, let us define the conditional Hamiltonian $G_{\tau}$,
\begin{equation} 
G_{\tau} \equiv \sum_{i=1}^{d}\bramatket{E_i}{U\ad H_{\tau} U}{E_i} U\dya{E_i}U\ad,
\end{equation}
where $E_i$ is the eigenenergy of the initial Hamiltonian $H_0$ with its corresponding eigenstate $\ket{E_i}$. At $t=0$ we perform {a} projective energy measurement $H_0$ on the system initially prepared in  {$\rho_0^{\eq}$}. Then, the post-measurement state will be projected onto $\ket{E_i}$ with the corresponding energy  $E_i$. After the evolution, the state {is} $U\ket{E_i}$. At $t=\tau$ we perform the second measurement $G_\tau$. Since the final state is a \textit{pointer state} of $G_\tau$, it {is not} destroyed by the measurement, so that the observer obtains the final energy measurement outcome $\bramatket{E_i}{U\ad H_{\tau} U}{E_i}$, while the system remains as $U\ket{E_i}$. {The corresponding quantum work is simply given by} Eq.~\eqref{eq:ConditionalWork}.

Then, {the equivalent work distribution within the TTM paradigm}, the forward work distribution is computed as
\begin{equation}
\widetilde{P}_f(W) = \sum_{i=1}^{d} \frac{e^{-\beta \bramatket{E_i}{H_{0}}{E_i}}}{Z_0}\abs{\bramatket{E_i}{U\ad U}{E_i}}^2\delta(W-\widetilde{W}_i),
\end{equation}
which is identical to Eq.~\eqref{eq:ForwardW}. 

\begin{figure}[htp!]
\centering
\includegraphics[width=1\columnwidth]{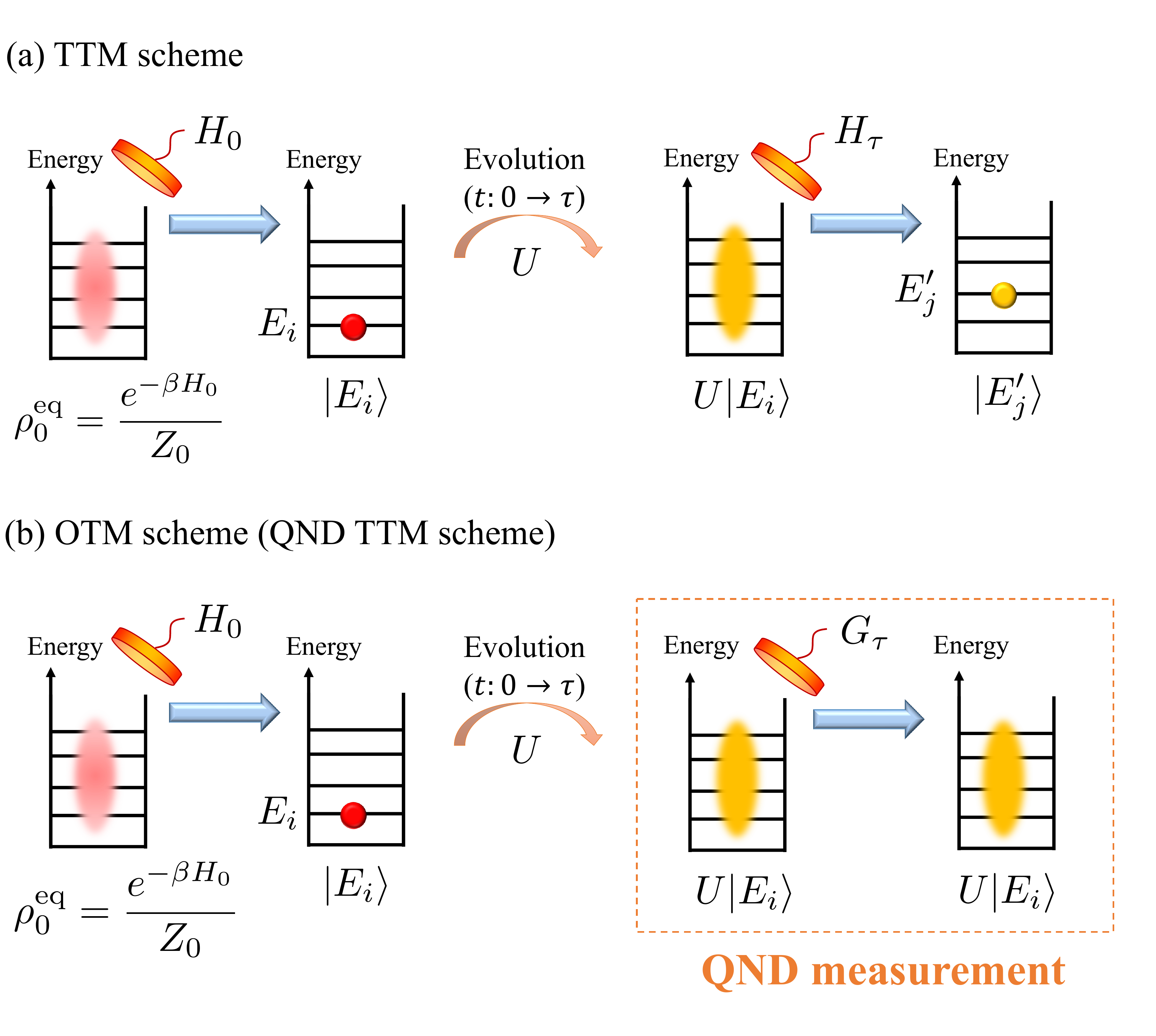}
\caption{Comparison between the TTM and OTM scheme. In (a) the standard TTM scheme, in which the second energy measurement is the final Hamiltonian, the projection measurement {projects} the evolved state $U\ket{E_i}$ onto $\ket{E_j'}$ the eigenstate of $H_{\tau}$. In (b) the OTM scheme, the final Hamiltonian $G_{\tau}\equiv \sum_{i=1}^{d} \bramatket{E_i}{U\ad H_{\tau}U}{E_i}U\dya{E_i}U\ad$ is the conditional Hamiltonian with its pointer state $U\ket{E_i}$, which is equivalent to the evolved state conditioned on the initial measurement outcome. Therefore, this measurement {preserves} the state $U\ket{E_i}$. In this sense, this measurement is  {a QND measurement}.} 
\label{fig:TTMvsOTM}
\end{figure}

{This is our \textit{first main result}, namely we have} that the OTM scheme is exactly the  {QND} TTM scheme, in which the second projection measurement does not destroy the evolved state conditioned on the initial measurement outcome~ (see Fig.~\ref{fig:TTMvsOTM} and \cite{supp}). {We will now exploit this insight to construct the conditional work distribution for the backward process within the OTM paradigm.}

{The backward process is initialized} from the state $\widetilde{\rho}_{\tau} 
\equiv {\exp{\left(-\beta G_{\tau}\right)}}/{\widetilde{Z}_{\tau}}$. After the backward evolution described by $U\ad$, the measurement $H_0$ is performed on the final state of the backward process. Since the final state is $U\ad U\ket{E_i}=\ket{E_i}$, which is a pointer state of $H_0$, $H_0$ does not destroy the state. Moreover, the outcome is always $E_i$. 

Then, by following the TTM scheme, the conditional work distribution of the backward process is given by 
\begin{equation}
\label{eq:BackW}
    \widetilde{P}_b(-W) \equiv\sum_{i=1}^{d}\frac{e^{-\beta\bramatket{E_i}{U\ad H_{\tau} U}{E_i}}}{\widetilde{Z}_{\tau}}\delta(-W+\widetilde{W}_i).
\end{equation}
{From Eq.~\eqref{eq:BackW} we now derive the fluctuation theorem}~\cite{Campisi11,mazzola2013measuring,dorner2013extracting} between the forward conditional work distribution and backward distribution in the characteristic function form. 

The characteristic functions are defined as the Fourier transform of the work distributions 
\begin{equation}
\begin{split}
\widetilde{C}_f(u) &\equiv \int dW \widetilde{P}_f(W) e^{iuW}\\
\widetilde{C}_b(u) &\equiv \int dW \widetilde{P}_b(-W) e^{-iuW}.
\end{split}
\end{equation}
By applying the approach to {the} TTM scheme in Ref.~\cite{Campisi11,mazzola2013measuring,dorner2013extracting}, the characteristic functions are equivalent to
\begin{equation}
\begin{split}
    \widetilde{C}_f(u) &= \tr{U\ad e^{iuG_{\tau}} U e^{-iuH_0}\rho^{\eq}_0}\\
    \widetilde{C}_b(u) &= \tr{U e^{iuH_0} U\ad e^{-iuG_{\tau}}\widetilde{\rho}_\tau}.
\end{split}
\label{eq:CharacFunction}
\end{equation}
Thus, we obtain the following {symmetry} relation~\footnote{In Supplemental Material~\cite{supp}, we provide the proof for the case that the initial state is also the conditional thermal state defined by $G_0\equiv\sum_{i=1}^{d}\bramatket{\psi_i}{H_0}{\psi_i}\dya{\psi_i}$ with $\ket{\psi_i}$ being not necessarily the eigenstate of $H_0$. Equation.~\eqref{eq:Main1} can be regarded as its corollary.
}, which is our \textit{second main result}, 
\begin{equation}
\frac{\widetilde{C}_f(u)}{\widetilde{C}_b(-u+i\beta)} = \frac{\widetilde{Z}_{\tau}}{Z_0} = e^{-\beta\Delta F-S(\widetilde{\rho}_{\tau}||\rho_{\tau}^{\eq})},
\label{eq:Main1}
\end{equation}
where 
\begin{equation}
\widetilde{C}_b(-u+i\beta)=\tr{U e^{-iuH_0}e^{-\beta H_0} U\ad e^{iuG_{\tau}}e^{\beta G_{\tau}}\widetilde{\rho}_\tau}.
\label{eq:BackWardCharacFunc}
\end{equation}
{The characteristic functions can be determined directly from quantum circuits, and hence our results permit the demonstration of} the experimental implementability of the OTM scheme in the single qubit interferometry.

\subparagraph{Single-qubit interferometry approach.}
By employing the single-qubit interferometry approach developed in Refs.~\cite{mazzola2013measuring,dorner2013extracting}, we {now} construct a quantum algorithm to {verify} Eq.~\eqref{eq:Main1}. This indicates that the OTM scheme {is} experimentally implementable, which is our \textit{third main result}. 

\begin{figure}
\mbox{
\Qcircuit @C=1em @R=1.2em {
\lstick{\ket{0}}& \gate{\mathbf{H}}& \ctrl{0} \qwx[1] &
\qw & \ctrl{0} \qwx[1] & \meter&~~~\text{$X$, $Y$} \\
\lstick{\rho_0^{\eq}}&\qw & \gate{e^{-i u H_0}} & \gate{U}& \gate{e^{iu G_{\tau}}} &\qw \\
}
}
\caption{Quantum circuit for computing $\widetilde{C}_f(u)$. The expectation values of $X$ and $Y$ obtained by measuring the final state of the ancilla qubit are $\Re[\widetilde{C}_f(u)]$ and $\Im[\widetilde{C}_f(u)]$, respectively.  }
\label{fig:ForwardCircuit}
\end{figure}
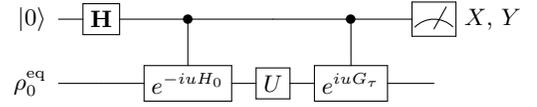

Let us define $\ket{0}\equiv\begin{pmatrix}1&0\end{pmatrix}^T$ and $\ket{1}\equiv\begin{pmatrix}0&1\end{pmatrix}^T$. We {denote by $\id$} the $2\times 2$ identity matrix and $X$, $Y$, $Z$ as the usual Pauli matrices. Also, we write the Hadamard gate as $\mathbf{H}\equiv\frac{1}{\sqrt{2}}\begin{pmatrix}1&1\\
1&-1\end{pmatrix}$. Then, the characteristic function of the forward process $\widetilde{C}_f(u)$ can be computed by the quantum circuit {depicted in Fig.~\ref{fig:ForwardCircuit}.} In this circuit, the ancilla qubit is initially prepared in $\ket{0}$. The target system is prepared in the Gibbs state $\rho_0^{\eq}$. To obtain the characteristic function $\widetilde{C}_f(u)$, we measure the output state of the ancilla qubit with $X$ and $Y$, whose expectation values become  $\ave{X}=\Re[\widetilde{C}_f(u)]$ and $\ave{Y}=\Im[\widetilde{C}_f(u)]$.

\begin{figure}[htp!]
\mbox{
\Qcircuit @C=1em @R=1.2em {
\lstick{\ket{0}}& \gate{\mathbf{H}}& \ctrl{0} \qwx[1] &
\ctrl{0} \qwx[1] & \qw & \ctrl{0} \qwx[1] & \ctrl{0} \qwx[1] &  \meter&~~~\text{$X$, $Y$} \\
\lstick{\widetilde{\rho}_{\tau}}&\qw & \gate{e^{iu G_{\tau}}}&\gate{\sigma_{\ell}} & \gate{U\ad}& \gate{e^{-iuH_{0}}}& \gate{\sigma_k} &\qw \\
}
}
\caption{\textbf{Quantum circuit for computing $F_{k\ell}$}: The expectation values of $X$ and $Y$ obtained by measuring the final state of the ancilla qubit are $\Re[F_{k\ell}]$ and $\Im[F_{k\ell}]$, respectively.}
\label{fig:BackwardCircuit}
\end{figure}
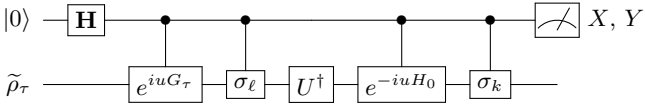

Next, {we} consider the quantum circuit to compute the characteristic function of the backward process $\widetilde{C}_b(-u+i\beta)$. From Eq.~\eqref{eq:BackWardCharacFunc}, we need to first decompose $\exp(\beta G_{\tau})$ and $\exp(-\beta H_0)$ with Pauli string $\{\sigma_k\}_{k=1}^{d^2}$, where $\sigma_1$ is the $d\times d$ identity matrix~\footnote{When $d=4$, the Pauli string is $\{\id, X\otimes \id, Y\otimes\id, Z\otimes\id, \id\otimes X, \id\otimes Y, \id\otimes Z, X\otimes X, X\otimes Y, X\otimes Z, Y\otimes X, Y\otimes Y, Y\otimes Z, Z\otimes X, Z\otimes Y, Z\otimes Z \}$, which has $4^2=16$ elements, and these elements are the bases constructing any $4\times 4$ matrix.}. Here, note that $\frac{1}{d}\tr{\sigma_k\sigma_\ell}=\delta_{k\ell}$ becomes Kronecker's delta. Then, we can write $\exp(-\beta H_{0}) = \sum_{k=1}^{d^2}\alpha_k^{(0)}\sigma_k$ and $\exp(\beta G_{\tau}) = \sum_{k=1}^{d^2}\alpha_k^{(\tau)}\sigma_k$, where $\alpha_k^{(0)}= \frac{1}{d} \tr{e^{-\beta H_0}\sigma_k}$ and $\alpha_k^{(\tau)}=\frac{1}{d} \tr{e^{\beta G_{\tau}}\sigma_k}$. Note that the coefficients $\{\alpha_k^{(0)},\alpha_k^{(\tau)}\}_{k=1}^{d^2}$ are computable via {a} classical computer {since} we have full knowledge of $H_0$, $H_{\tau}$, and $U$ if $d$ is smaller. Therefore, we can write 
\begin{equation}
    \widetilde{C}_b(-u+i\beta)=\sum_{k,\ell}\alpha_k^{(0)}\alpha_\ell^{(\tau)}F_{k\ell},
\label{eq:CbandFvalue}
\end{equation}
where we define
\begin{equation}
    F_{k\ell}\equiv \tr{U\sigma_k e^{-i u H_0}U\ad \sigma_\ell e^{iu G_{\tau}}\widetilde{\rho}_{\tau}}.
\label{eq:Fvalue}
\end{equation}
Then, we can employ the quantum circuit {depicted in Fig.~\ref{fig:BackwardCircuit}} to compute $F_{k\ell}$. In this circuit, the ancilla qubit is prepared in $\ket{0}$. The target system is prepared in the conditional thermal state $\widetilde{\rho}_{\tau}$. Here, the expectation values become $\ave{X}=\Re[F_{k\ell}]$ and $\ave{Y}=\Im[F_{k\ell}]$. Given the fact that we have already known $\{\alpha_{k}^{(0)}, \alpha_{k}^{(\tau)}\}_{k=1}^{d^2}$ via {a} classical computer, from Eq.~\eqref{eq:CbandFvalue}, we can finally obtain $\widetilde{C}_b(-u+i\beta)$.

\subparagraph{ {Physical meaning of OTM fluctuation theorem.}}

 {By considering the backward process, we can find that the OTM fluctuation theorem can capture the detailed information about the irreversibility and be applied to state preparation for quantum thermometry in the low-temperature limit. To quantify the irreversibility of a quantum process, we consider the Kullback-Leibler (KL) divergence $D[\widetilde{P}_f||\widetilde{P}_b]$ of $\widetilde{P}_f(W)$ with respect to $\widetilde{P}_b(-W)$, which is defined as 
\begin{align}
D[\widetilde{P}_f||\widetilde{P}_b]\equiv \int dW \widetilde{P}_f(W)\ln\left(\frac{\widetilde{P}_f(W)}{\widetilde{P}_b(-W)}\right).
\end{align}
From Eqs.~\eqref{eq:ForwardW} and \eqref{eq:BackW}, we obtain~\cite{supp} 
\begin{align}
\!\!\!\!D[\widetilde{P}_f||\widetilde{P}_b]\!=\!-S(\rho_0^{\eq})+\beta\tr{U\rho_0^{\eq}U\ad H_{\tau}}+\ln\widetilde{Z}_{\tau},   
\label{eq:KLdivergence}
\end{align}
where $S(\rho_0^{\eq})\equiv-\tr{\rho_0^{\eq}\ln\rho_0^{\eq}}$ is the von-Neumann entropy of $\rho_0^{\eq}$. Given that the exact averaged work $\ave{W}\equiv\tr{U\rho_0^{\eq}U\ad H_{\tau}}-\tr{\rho_0^{\eq}H_0}$, from Eq.~\eqref{eq:Main1}, the excess work $\ave{W_{\text{ex}}}\equiv\ave{W}-\Delta F$ can be written as 
\begin{align}
    \beta\ave{W_{\text{ex}}} = D[\widetilde{P}_f||\widetilde{P}_b]+S(\widetilde{\rho}_{\tau}||\rho_{\tau}^{\eq}).
\label{eq:exW}
\end{align}
This means that the excess work $\ave{W_{\text{ex}}}$ is a sum of the KL divergence  $D[\widetilde{P}_f||\widetilde{P}_b]$, which characterizes the irreversible process, and $\beta S(\widetilde{\rho}_{\tau}||\rho_{\tau}^{\eq})$, which is the energy dissipated into the heat bath when the system is thermalized from $\widetilde{\rho}_{\tau}$. Therefore, {$\beta S(\widetilde{\rho}_{\tau}||\rho_{\tau}^{\eq})$} can be interpreted as a heatlike quantity. }

 {Furthermore, $D[\widetilde{P}_f||\widetilde{P}_b]$ can be employed in quantum thermometry in the low-temperature limit. In Ref.~\cite{sone2023conditional}, it was demonstrated that the conditional thermal state $\widetilde{\rho}_{\tau}$ can outperform the Gibbs state $\rho_{\tau}^{\eq}$ in the low-temperature limit. Therefore, preparing $\widetilde{\rho}_{\tau}$ is a desired task for quantum thermometry.  First the excess work can be written as 
$\beta\ave{W_{\text{ex}}}=S(U\rho_0^{\eq}U\ad||\rho_{\tau}^{\eq})$~\cite{deffner2010generalized,kawai2007dissipation,vaikuntanathan2009dissipation}. In Ref.~\cite{sone2023conditional}, we derived the so-called thermodynamic triangle equality $S(U\rho_0^{\eq}U\ad||\widetilde{\rho}_{\tau})+S(\widetilde{\rho}_{\tau}||\rho_{\tau}^{\eq})=S(U\rho_0^{\eq}U\ad||\rho_{\tau}^{\eq})$. Therefore, from Eq.~\eqref{eq:exW}, we obtain
\begin{align}
   D[\widetilde{P}_f||\widetilde{P}_b]=S(U\rho_0^{\eq}U\ad||\widetilde{\rho}_{\tau}),
\end{align}
which measures the distinguishability of the exact final state $U\rho_0^{\eq}U\ad$ and the conditional thermal state $\widetilde{\rho}_{\tau}$. This quantity can be used to design the unitary process $U$ that minimizes $S(U\rho_0^{\eq}U\ad||\widetilde{\rho}_{\tau})$ for the final exact state $U\rho_0^{\eq}U\ad$ to be closer to $\widetilde{\rho}_{\tau}$. Also, note that the distinguishability measure $S(U\rho_0^{\eq}U\ad||\widetilde{\rho}_{\tau})$ can be computed by a quantum computer. From Eqs.~\eqref{eq:Main1} and \eqref{eq:KLdivergence}, we have
\begin{align}
D[\widetilde{P}_f||\widetilde{P}_b] = \beta\ave{W}+\ln\left(\frac{\widetilde{C}_f(u)}{\widetilde{C}_b(-u+i\beta)}\right),
\end{align}
where $\ave{W}$, $\widetilde{C}_f(u)$, and $\widetilde{C}_b(-u+i\beta)$ can be computed by a quantum computer.}

 {Finally, we emphasize that these analyses are hard to conduct within the TTM scheme~\cite{kiely2023entropy}. Therefore, our detailed fluctuation demonstrates an additional advantage of the OTM scheme. }

\subparagraph{{Verification with IBM quantum computers}}

{To conclude our analysis, we } employ the IBM cloud-based quantum computer~\cite{cross2018ibm} to verify the {detailed fluctuation theorem}~\eqref{eq:Main1}. Our setup is the following. The initial Hamiltonian $H_0$ is $H_0 = \omega (Z\otimes\id + \id\otimes Z)$
with the corresponding eigenbasis
$\ket{E_1} = \begin{pmatrix}1&0&0&0\end{pmatrix}^{T}$, $\ket{E_2} = \begin{pmatrix}0&1&0&0\end{pmatrix}^{T}$, $\ket{E_3} = \begin{pmatrix}0&0&1&0\end{pmatrix}^{T}$, and $\ket{E_4} = \begin{pmatrix}0&0&0&1\end{pmatrix}^{T}$. The final Hamiltonian is set to be $H_\tau = J (X \otimes X)$. The unitary operator that describes the evolution is set as $U = \exp\left(-i \frac{\Omega \tau}{2}(Y \otimes \id +\id \otimes Y)\right)$. 

For the initial Gibbs state preparation, we consider the decomposition of the input mixed state. This is because we can only prepare {pure states on} the IBM quantum computers {and} $\{\ket{E_i}\}_{i=1}^{4}$ can be prepared. For the weights $\{\exp(-\beta E_i)/Z_0\}_{i=1}^{4}$, {since we already know} the initial Hamiltonian $H_0$, we assume that the weights are {also known and we can} compute $\widetilde{C}_f(u)$. 

For the backward process, we need to prepare the conditional thermal state $\widetilde{\rho}_{\tau}$ as the initial state. In our {simulation}, we suppose that we {already know} $U$; therefore, we can prepare $\{U\ket{E_i}\}_{i=1}^{4}$ with the quantum computer. Similarly, because we assume that $U$, $H_{0}$ and $H_{\tau}$ are known, the weights $\{\exp\left(-\beta\bramatket{E_i}{U\ad H_{\tau} U}{E_i}\right)/\widetilde{Z}_{\tau}\}_{i=1}^{4}$ are considered to be {also} known, which we use to compute $F_{k\ell}$. {Thus} the coefficients $\alpha_k^{(0)}$ and $\alpha_{\ell}^{(\tau)}$ are also regarded as known values, which enables one to compute $\widetilde{C}_b(-u+i\beta)$.

By setting the parameters as $\beta=0.5$, $\omega=2$, $\Omega=3$, $J=1$, and $\tau=\pi/4$, we obtain the theoretical value of the ratio 
\begin{equation}
R_{\text{true}}\equiv\frac{\widetilde{C}_f(u)}{\widetilde{C}_b(-u+i\beta)}=0.433167
\label{eq:ExactValue}
\end{equation}
for any $u$. Here, we particularly focus on the case $u = 1$, and verify Eq.~\eqref{eq:ExactValue} with the IBM cloud-based quantum computer~\cite{cross2018ibm}.

To {determine} $\avg{X}$ and $\avg{Y}$, for each quantum circuit in Figs.~\ref{fig:ForwardCircuit} and \ref{fig:BackwardCircuit}, we perform the single-shot measurement $20000$ times. The median of the errors of the gates and the single-qubit readout error in our setup are around $10^{-4}\sim 10^{-2}$ with the $T_1$ and $T_2$ ranging from around 17 $\upmu\text{s} \sim 232 \upmu\text{s}$~(for complete information of the IBM machine, refer to the Supplemental Material~\cite{supp}). Because of the error, the computed ratio becomes complex; therefore, we consider the absolute value of the ratio $ R\equiv \abs{\widetilde{C}_f(1)/\widetilde{C}_b(-1+0.5i)}$. To obtain more precise {values}, we run the whole process $N$ times (number of trials) and compare the true value with the average value $\ave{R}_N\equiv\frac{1}{N}\sum_{j=1}^{N}R_j$, where $R_j$ is the value of $R$ at the $j$th trial. Then, we compute the error rate as 
$e_N=\abs{1-\ave{R}_N/R_{\text{true}}}\times 100~[\%]$ for each $N$. 

\begin{figure}
\centering
\includegraphics[width=1\columnwidth]{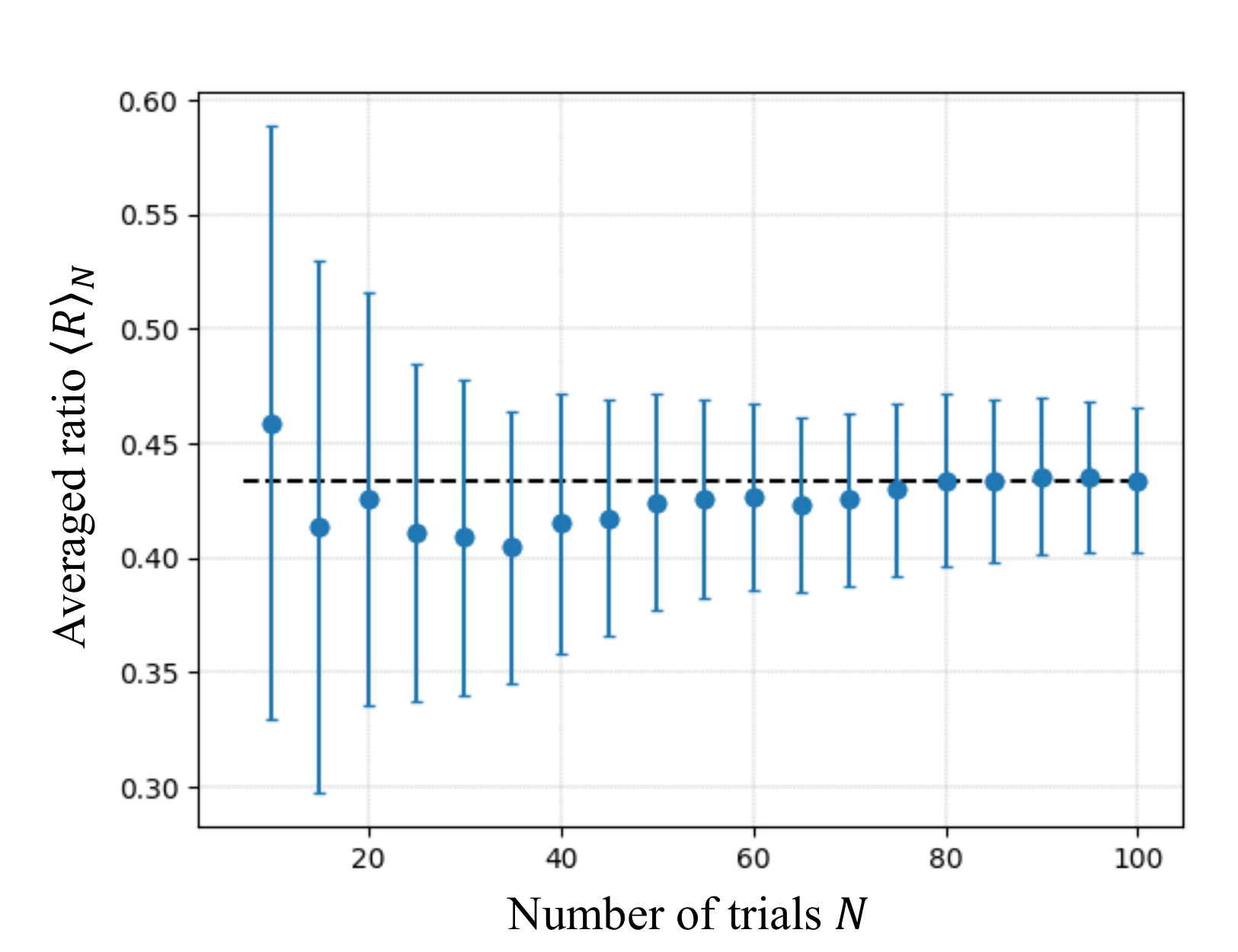}
\caption{Verification with IBM quantum computers. The dashed line is the exact value $R_{\text{true}}=0.433167$. The error bars represent $99\%$ confidence interval. As we increase the number of trials $N$, the averaged ratios $\ave{R}_N$ approach the exact value with a clear plateau starting from $N=75$. When $N=100$, we have $\ave{R}_{100}=0.433706$. }
\label{fig:ExpResult}
\end{figure}

We have achieved a very high accuracy in our {simulation}. In Fig.~\ref{fig:ExpResult}, we plot the relation between $\ave{R}_N$ for each number of trials $N=10,15,20,\cdots,100$, where the error bars represent the 99\% confidence interval~\cite{dekking2005modern}. As we can see, as we increase the number of trials, the average value converges to the true value with explicit plateau starting from $N=75$. When $N=100$, $\ave{R}_{100}$ records 0.433706. 

\begin{figure}
\centering
\includegraphics[width=1\columnwidth]{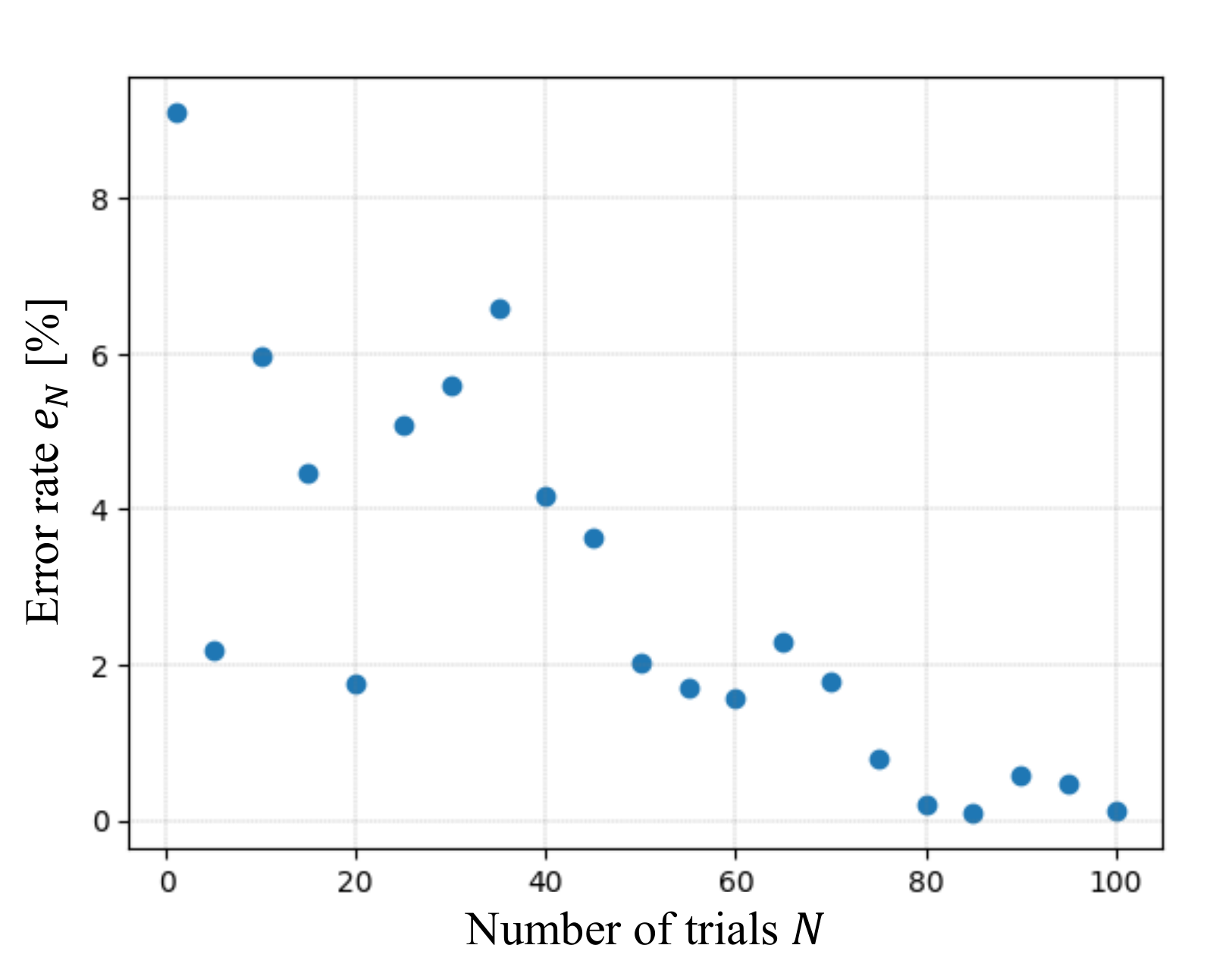}
\caption{Error rate vs number of trials. As we increase the number of trials $N$, the error rate $e_N$ becomes smaller. When $N=100$, we have $e_{100}\simeq0.12\%$. }
\label{fig:ErrorResult}
\end{figure}

In Fig.~\ref{fig:ErrorResult}, we show the relation between the error rate $e_N$ and the number of trials $N$. As we can see, as the number of trials increases, the error rate becomes smaller. Actually, when $N=100$, $e_{100}$ records around $0.12\%$, which is accurate enough to claim that the OTM fluctuation theorem Eq.~\eqref{eq:Main1} is verified with the IBM quantum computer.

\subparagraph{Conclusion.}
In conclusion, we have derived the detailed fluctuation theorem of the OTM scheme by clarifying the backward work distribution. {This has been enabled by the insight} that the OTM scheme {can be regarded as} a  {QND} TTM scheme, where the second measurement is constructed by the pointer states conditioned on the initial energy measurement outcome.  {We have related its physical meaning to the irreversibility and quantum thermometry in the low-temperature limit.} We have also demonstrated its experimental implementability {and we have verified} the derived fluctuation theorem on the IBM quantum computer by introducing the corresponding quantum circuit to compute the symmetric relation between the characteristic functions of the forward and backward work distributions. These results not only provide the solutions to the open problems regarding the OTM scheme, but also clarify that the laws of thermodynamics {at} the nanoscale are strictly dependent on the choice of the measurement of the observer. From {a} practical point of view, these results provide  experimentalists with a concrete strategy to study laws of thermodynamics {at} the nanoscale by protecting quantum coherence and correlations.

\begin{acknowledgements}
 {This work is supported by the NSF under Grant No. MPS-2328774.}  {A.S. is} grateful to S. Endo for helpful discussions. K.M. is supported by the Goldwater scholarship. T.H. is supported by the graduate study program at the University of Massachusetts Boston. 
{S.D. acknowledges support from  the U.S. National Science Foundation under Grant No. DMR-2010127 and the John Templeton Foundation under Grant No. 62422.}
\end{acknowledgements}

\bibliography{ref.bib}

\newpage

\onecolumngrid

\setcounter{equation}{0}
\setcounter{figure}{0}
\renewcommand{\theequation}{S\arabic{equation}}
\renewcommand{\thefigure}{S\arabic{figure}}
\renewcommand{\thetable}{S\arabic{table}}

\appendix

\section*{\large{Supplementary Material for \\``Detailed Fluctuation Theorem from the One-Time Measurement Scheme'' }}

\section{{OTM scheme as a QND TTM scheme}}

{In the OTM scheme, focusing on the forward process, when the final measurement is 
\begin{align}
G_{\tau} \equiv \sum_{i=1}^{d}\bramatket{E_i}{U\ad G_{\tau} U}{E_i} U\dya{E_i}U\ad
\end{align}
and the state before the final measurement is $U\ket{E_i}$, the state does not change due to $G_{\tau}$ as~\cite{Nielsen}  
\begin{align}
\frac{G_{\tau} U\ket{E_i}}{\sqrt{\bramatket{E_i}{U\ad G_{\tau}\ad G_{\tau} U}{E_i}}}= U\ket{E_i}.
\end{align}
This demonstrates that $G_{\tau}$ is a QND measurement, so that OTM scheme can be classified as the QND TTM scheme.}

\section{Derivation of the detailed fluctuation theorem and the symmetric relation}

To generalize our formalism, we consider the case that the initial state is also a conditional thermal state 
\begin{equation}
\widetilde{\rho}_0 \equiv \sum_{i=1}^{d}\frac{e^{-\beta \bramatket{\psi_i}{H_{0}}{\psi_i}}}{\widetilde{Z}_{0}}\dya{\psi_i}.
\end{equation}
Here, note that $\ket{\psi_i}$ is not necessarily the eigenstate of $H_0$. Defining 
\begin{equation}
G_0\equiv\sum_{i=1}^{d}\bramatket{\psi_i}{H_0}{\psi_i}\dya{\psi_i},
\end{equation}
where $\bramatket{\psi_i}{H_0}{\psi_i}$ and $\ket{\psi_i}$ are the eigenvalue and the corresponding eigenstate of $G_0$, we can write
\begin{equation}
    \widetilde{\rho}_0 = \frac{e^{-\beta G_0}}{\widetilde{Z}_0}
\end{equation}
with $\widetilde{Z}_0\equiv\tr{\exp(-\beta G_0)}$ the normalization factor. Here, note that when $\ket{\psi_i}=\ket{E_i}$, we have $G_0=H_0$ and $\widetilde{\rho}_0=\rho_0^{\eq}$. 

We first consider the following TTM scheme. At time $t=0$, we measure the system with $G_0$. Suppose that the outcome is $\bramatket{\psi_i}{H_{0}}{\psi_i}$; therefore, the post-measurement state will be projected onto $\ket{\psi_i}$. 
After the evolution $U$, we obtain a final state $U\ket{\psi_i}$. At time $t=\tau$ we perform a measurement $G_{\tau}$ defined as 
\begin{equation}
G_{\tau} \equiv \sum_{i=1}^{d}\bramatket{\psi_i}{U\ad H_{\tau} U}{\psi_i} U\dya{\psi_i}U\ad.
\label{eq:app:Gtau}
\end{equation}
Here, note that $G_{\tau}$ does not destroy the final state $U\ket{\psi_i}$ because $U\ket{\psi_i}$ itself is the pointer state of $G_{\tau}$. Therefore, the outcome of $G_{\tau}$ is always $\bramatket{\psi_i}{U\ad H_{\tau} U}{\psi_i}$. We define
\begin{equation}
\widetilde{W}_i \equiv \bramatket{\psi_i}{U\ad H_{\tau} U}{\psi_i} - \bramatket{\psi_i}{H_{0}}{\psi_i},
\end{equation}
which is a random variable determined
in a single measurement trajectory. Then, applying the way of constructing the work distribution of the forward process in the TTM scheme, for the forward process, we can write
\begin{equation}
\begin{split}
\widetilde{P}_f(W) &= \sum_{i=1}^{d} \frac{e^{-\beta \bramatket{\psi_i}{H_{0}}{\psi_i}}}{\widetilde{Z}_{0}}\abs{\bramatket{\psi_i}{U\ad U}{\psi_i}}^2\delta(W-\widetilde{W}_i)\\
&=\sum_{i=1}^{d} \frac{e^{-\beta \bramatket{\psi_i}{H_{0}}{\psi_i}}}{\widetilde{Z}_{0}}\delta(W-\widetilde{W}_i).
\end{split}
\label{eq:app:Forward}
\end{equation}

For the backward process, we start from the state
\begin{equation}
\!\!\!\widetilde{\rho}_{\tau} 
\equiv \frac{e^{-\beta G_{\tau}}}{\widetilde{Z}_{\tau}}
= \sum_{i=1}^{d}\frac{e^{-\beta\bramatket{\psi_i}{U\ad H_{\tau} U}{\psi_i}}}{\widetilde{Z}_{\tau}}U\dya{\psi_i}U\ad.
\end{equation}
Then, after the backward evolution described by $U\ad$, we perform the measurement $G_0$ on the final state of the backward process. Again, since the final state is $U\ad U\ket{\psi_i}=\ket{\psi_i}$, which is the pointer state of $G_0$, $G_0$ does not destroy the state $\ket{\psi_i}$. Therefore, the outcome is always $\bramatket{\psi_i}{H_{0}}{\psi_i}$. Then, the work distribution of the backward process is given by 
\begin{equation}
\begin{split}
\widetilde{P}_b(-W) \equiv& \sum_{i=1}^{d}\frac{e^{-\beta\bramatket{\psi_i}{U\ad H_{\tau} U}{\psi_i}}}{\widetilde{Z}_{\tau}}\abs{\bramatket{\psi_i}{U\ad U}{\psi_i}}^2\delta(-W+\widetilde{W}_i)\\
=&\sum_{i=1}^{d}\frac{e^{-\beta\bramatket{\psi_i}{U\ad H_{\tau} U}{\psi_i}}}{\widetilde{Z}_{\tau}}\delta(-W+\widetilde{W}_i).
\end{split}
\label{eq:app:Backward}
\end{equation}

{Since}
\begin{equation}
\begin{split}
\widetilde{P}_f(W) 
&=\sum_{i=1}^{d} \frac{e^{-\beta \bramatket{\psi_i}{H_{0}}{\psi_i}}}{\widetilde{Z}_{0}}\delta(W-\widetilde{W}_i)\\
& = \frac{\widetilde{Z}_{\tau}}{\widetilde{Z}_{0}}\sum_{i=1}^{d} \frac{e^{\beta (W-\bramatket{\psi_i}{U\ad H_{\tau} U}{\psi_i})}}{\widetilde{Z}_{\tau}}\delta(-W+\widetilde{W}_i)\\
&=\frac{\widetilde{Z}_{\tau}}{\widetilde{Z}_{0}}e^{\beta W}\widetilde{P}_b(-W),
\end{split}
\end{equation}
which yields 
\begin{equation}
\frac{\widetilde{P}_f(W)}{\widetilde{P}_b(-W)}e^{-\beta W} = \frac{\widetilde{Z}_{\tau}}{\widetilde{Z}_{0}}.
\end{equation}

Let us compute the following quantum relative entropies $S(\widetilde{\rho}_{\tau}||\rho_{\tau}^{\eq})$ and $S(\widetilde{\rho}_0||\rho_0^{\eq})$. From Ref.~\cite{Deffner16}, it has already been proven that
\begin{equation}
S(\widetilde{\rho}_{\tau}||\rho_{\tau}^{\eq}) = -\ln\left(\frac{\widetilde{Z}_{\tau}}{Z_{\tau}}\right).
\end{equation}
Similarly, for $S(\widetilde{\rho}_0||\rho_0^{\eq})$, we have
\begin{equation}
S(\widetilde{\rho}_{0}||\rho_{0}^{\eq}) = -\ln\left(\frac{\widetilde{Z}_{0}}{Z_{0}}\right).
\end{equation}
{Since} the equilibrium Helmholtz free energy difference is given by
\begin{equation}
\Delta F = -\frac{1}{\beta}\ln\left(\frac{Z_{\tau}}{Z_0}\right),
\end{equation}
we can write
\begin{equation}
S(\widetilde{\rho}_{\tau}||\rho_{\tau}^{\eq})-
S(\widetilde{\rho}_{0}||\rho_{0}^{\eq}) = \ln\left(\frac{Z_{\tau}}{Z_0}\right)-\ln\left(\frac{\widetilde{Z}_{\tau}}{\widetilde{Z}_0}\right)
= -\beta\Delta F-\ln\left(\frac{\widetilde{Z}_{\tau}}{\widetilde{Z}_{0}}\right).
\label{eq:app:RelativeEntropyDifference}
\end{equation}
Therefore, we can obtain the following detailed fluctuation theorem
\begin{equation}
\frac{\widetilde{P}_f(W)}{\widetilde{P}_b(-W)}e^{-\beta W} = e^{-\beta \Delta F-S(\widetilde{\rho}_{\tau}||\rho_{\tau}^{\eq})+S(\widetilde{\rho}_{0}||\rho_{0}^{\eq})}.
\end{equation}
Note that $\widetilde{P}_f(W)$ and $\widetilde{P}_b(-W)$ cross at $W=\Delta F+\beta^{-1}(S(\widetilde{\rho}_{\tau}||\rho_{\tau}^{\eq})-S(\widetilde{\rho}_{0}||\rho_{0}^{\eq}))$. 
Also, in the setup in Ref.~\cite{Deffner16} (i.e. $\ket{\psi_i}=\ket{E_i}$), we have  $\widetilde{\rho}_0=\rho_0^{\eq}$ so that $S(\widetilde{\rho}_0||\rho_0^{\eq})=0$. Therefore, we can recover the OTM fluctuation theorem and the corresponding Jarzynski equality in Ref.~\cite{Deffner16}.

Next, we derive the symmetric relation by considering the characteristic functions of the work distributions~\cite{Campisi11,mazzola2013measuring,dorner2013extracting}. The characteristic functions are defined as 
\begin{equation}
\begin{split}
\widetilde{C}_f(u) &\equiv \int dW \widetilde{P}_f(W) e^{iuW}\\
\widetilde{C}_b(u) &\equiv \int dW \widetilde{P}_b(-W) e^{-iuW}.
\end{split}
\end{equation}
From Eq.~\eqref{eq:app:Forward}, we have 
\begin{equation}
\begin{split}
\widetilde{C}_f(u) &= \sum_{i=1}^{d}\int dW\frac{e^{-\beta\bramatket{\psi_i}{H_{0}}{\psi_i}}}{\widetilde{Z}_0}e^{iuW}\delta(W-\widetilde{W}_i)\\
&=\sum_{i=1}^{d}\frac{e^{-\beta\bramatket{\psi_i}{H_{0}}{\psi_i}}}{\widetilde{Z}_0} e^{iu\widetilde{W}_i}\\
&=\sum_{i=1}^{d}\frac{e^{-(iu+\beta)\bramatket{\psi_i}{H_{0}}{\psi_i}}}{\widetilde{Z}_0}e^{iu\bramatket{\psi_i}{U\ad H_{\tau} U}{\psi_i}}\\
&= \frac{\widetilde{Z}_{\tau}}{\widetilde{Z}_{0}} \sum_{i=1}^{d}\frac{e^{iu\bramatket{\psi_i}{U\ad H_{\tau} U}{\psi_i}}}{\widetilde{Z}_{\tau}} e^{-(iu+\beta)\bramatket{\psi_i}{H_{0}}{\psi_i}}.
\end{split}
\label{eq:app:ForwardCharac}
\end{equation}
Furthermore, from Eq.~\eqref{eq:app:Backward} and $\delta(-W+\widetilde{W_i})=\delta(W-\widetilde{W_i})$, we can obtain 
\begin{equation}
\begin{split}
\widetilde{C}_b(-u+i\beta)=&
\sum_{i=1}^{d}\int dW \frac{e^{-\beta\bramatket{\psi_i}{U\ad H_{\tau} U}{\psi_i}}}{\widetilde{Z}_{\tau}}e^{-i(-u+i\beta)W}\delta(-W+\widetilde{W_i})\\
=&\sum_{i=1}^{d}\int dW \frac{e^{-\beta\bramatket{\psi_i}{U\ad H_{\tau} U}{\psi_i}}}{\widetilde{Z}_{\tau}}e^{(iu+\beta)\widetilde{W_i}}\\
=&\sum_{i=1}^{d}\frac{e^{iu\bramatket{\psi_i}{U\ad H_{\tau} U}{\psi_i}}}{\widetilde{Z}_{\tau}}e^{-(iu+\beta)\bramatket{\psi_i}{H_{0}}{\psi_i}}
\end{split}
\label{eq:app:BackwardCharac}
\end{equation}
Therefore, we can obtain the symmetric relation
\begin{equation}
\frac{\widetilde{C}_f(u)}{\widetilde{C}_b(-u+i\beta)} = \frac{\widetilde{Z}_{\tau}}{\widetilde{Z}_0}.
\label{eq:app:SymmetricRelation}
\end{equation}
From Eq.~\eqref{eq:app:RelativeEntropyDifference}, we can write
\begin{equation}
\frac{\widetilde{C}_f(u)}{\widetilde{C}_b(-u+i\beta)} = e^{-\beta\Delta F-S(\widetilde{\rho}_{\tau}||\rho_{\tau}^{\eq})+S(\widetilde{\rho}_0||\rho_0^{\eq})}.
\end{equation}

Equation.~\eqref{eq:app:SymmetricRelation} can be recovered from the quantum circuit representation. Following Refs.~\cite{Campisi11,mazzola2013measuring,dorner2013extracting}, the characteristic functions are
\begin{equation}
\begin{split}
\widetilde{C}_f(u)&=\tr{U\ad  e^{iuG_{\tau}} U e^{-iuG_0}\widetilde{\rho}_0}\\
\widetilde{C}_b(u)&=\tr{U e^{iuG_0} U\ad  e^{-iuG_{\tau}}\widetilde{\rho}_\tau}.
\end{split}
\end{equation}
From Eq.~\eqref{eq:app:Gtau}, we have
\begin{equation}
U\ad G_{\tau} U = \sum_{i=1}^{d}\bramatket{\psi_i}{U\ad H_{\tau} U}{\psi_i}\dya{\psi_i}.
\end{equation}
Therefore, 
\begin{equation}
\widetilde{C}_f(u) =\frac{\widetilde{Z}_{\tau}}{\widetilde{Z}_0} \sum_{i=1}^{d} \frac{e^{iu \bramatket{\psi_i}{U\ad H_{\tau} U}{\psi_i}}}{\widetilde{Z}_{\tau}}e^{-(iu+\beta)\bramatket{\psi_i}{H_0}{\psi_i}},
\end{equation}
which is equivalent to Eq.~\eqref{eq:app:ForwardCharac}.

Next, for the backward process, because
\begin{equation}
U G_0 U\ad = \sum_{i=1}^{d}\bramatket{\psi_i}{H_0}{\psi_i}U\dya{\psi_i}U\ad,
\end{equation}
we can obtain 
\begin{equation}
\widetilde{C}_b(u) = \sum_{i=1}^{d}\frac{e^{(-iu-\beta)\bramatket{\psi_i}{U\ad H_{\tau} U}{\psi_i}}}{\widetilde{Z}_{\tau}}e^{iu\bramatket{\psi_i}{H_0}{\psi_i}},
\end{equation}
so that we can arrive at \begin{equation}
\widetilde{C}_b(-u+i\beta) =
\sum_{i=1}^{d}\frac{e^{iu\bramatket{\psi_i}{U\ad H_{\tau} U}{\psi_i}}}{\widetilde{Z}_{\tau}}e^{-(iu+\beta)\bramatket{\psi_i}{H_{0}}{\psi_i}},
\end{equation}
which is equivalent to Eq.~\eqref{eq:app:BackwardCharac}. Hence, we can obtain
\begin{equation}
\begin{split}
\widetilde{C}_f(u) &=\tr{U\ad  e^{iuG_{\tau}} U e^{-iuG_0}\widetilde{\rho}_0}\\
&= \tr{U\ad  e^{iuG_{\tau}} U e^{-iuG_0}\frac{e^{-\beta G_0}}{\widetilde{Z}_{0}}}\\
&= \frac{1}{\widetilde{Z}_{0}}\tr{U\ad  e^{iuG_{\tau}} U e^{-iuG_0}e^{-\beta G_0}}\\
&= \frac{1}{\widetilde{Z}_{0}}\tr{U e^{-iuG_0}e^{-\beta G_0}U\ad  e^{iuG_{\tau}}}\\
&= \frac{1}{\widetilde{Z}_{0}}\tr{U e^{-iuG_0}e^{-\beta G_0}U\ad  e^{iuG_{\tau}}e^{\beta G_{\tau}}e^{-\beta G_{\tau}}}\\
&= \frac{\widetilde{Z}_{\tau}}{\widetilde{Z}_{0}}\tr{U e^{-iuG_0}e^{-\beta G_0}U\ad  e^{iuG_{\tau}}e^{\beta G_{\tau}}\frac{e^{-\beta G_{\tau}}}{\widetilde{Z}_{\tau}}}\\
&= \frac{\widetilde{Z}_{\tau}}{\widetilde{Z}_{0}}\tr{U e^{i(-u+i\beta)G_0}U\ad  e^{-i(-u+i\beta)G_{\tau}}\widetilde{\rho}_\tau}\\
&= \frac{\widetilde{Z}_{\tau}}{\widetilde{Z}_{0}}\widetilde{C}_b(-u+i\beta),
\end{split}
\end{equation}
which leads Eq.~\eqref{eq:app:SymmetricRelation}. Again, for the case $\ket{\psi_i}=\ket{E_i}$, because $\widetilde{\rho}_0=\rho_0^{\eq}$, we can recover our first main result $\widetilde{C}_f(u)/\widetilde{C}_b(-u+i\beta)  = \exp(-\beta\Delta F-S(\widetilde{\rho}_{\tau}||\rho_{\tau}^{\eq}))$.

\section{ { Derivation of $D[\widetilde{P}_f||\widetilde{P}_b]$}}

 {The KL divergence  $D[\widetilde{P}_f||\widetilde{P}_b]$ is defined as 
\begin{align}
D[\widetilde{P}_f||\widetilde{P}_b] \equiv \int dW \widetilde{P}_f(W)\ln\frac{\widetilde{P}_f(W)}{\widetilde{P}_b(-W)}.
\end{align}
Because
\begin{align}
    \widetilde{P}_f(W) &\equiv \sum_{i=1}^{d}\frac{e^{-\beta E_i}}{Z_0}\delta\left(W-\widetilde{W}_i\right)\\
    \widetilde{P}_b(-W) &\equiv\sum_{i=1}^{d}\frac{e^{-\beta\bramatket{E_i}{U\ad H_{\tau} U}{E_i}}}{\widetilde{Z}_{\tau}}\delta(-W+\widetilde{W}_i)
\end{align}
with
\begin{align}
    \widetilde{W}_i\equiv \bramatket{E_i}{U\ad H_{\tau} U}{E_i}-E_i,
\end{align}
we obtain 
\begin{align}
\begin{split}
D[\widetilde{P}_f||\widetilde{P}_b] =& \int dW \widetilde{P}_f(W)\ln\frac{\widetilde{P}_f(W)}{\widetilde{P}_b(-W)}\\
=&\int dW\sum_{i=1}^{d}\frac{e^{-\beta E_i}}{Z_0}\delta(W-\widetilde{W}_i)\ln\left(\sum_{j=1}^{d}\frac{e^{-\beta E_j}}{Z_0}\delta(W-\widetilde{W}_j)\right)\\
&-\int dW \sum_{i=1}^{d}\frac{e^{-\beta E_i}}{Z_0}\delta(W-\widetilde{W}_i)\ln\left(\sum_{j=1}^{d}\frac{e^{-\beta\bramatket{E_j}{U\ad H_{\tau} U}{E_j}}}{\widetilde{Z}_{\tau}}\delta(-W+\widetilde{W}_j)\right)\\
=&\sum_{i=1}^{d}\frac{e^{-\beta E_i}}{Z_0}\ln\left(\frac{e^{-\beta E_i}}{Z_0}\right)+\beta\left(\sum_{i=1}^{d}\frac{e^{-\beta E_i}}{Z_0}\bramatket{E_i}{U\ad H_{\tau} U}{E_i}\right)+\ln\widetilde{Z}_{\tau}\\
=&-S(\rho_0^{\eq})+\beta\tr{U\rho_0^{\eq}U\ad H_{\tau}}+\ln\widetilde{Z}_{\tau}\,,
\end{split}
\end{align}
where $S(\rho_0^{\eq})$ is the von-Neumann entropy of $\rho_0^{\eq}$.} 

\section{Detailed Qiskit circuits to compute the characteristic functions}
In this section, we provide the details of the quantum circuits to compute the characteristic functions $\widetilde{C}_f(1)$ and $\widetilde{C}_b(-1+0.5i)$.

\subsection{Forward process}
\begin{figure}[htp!]
  \centering
  \includegraphics[width=1\textwidth]{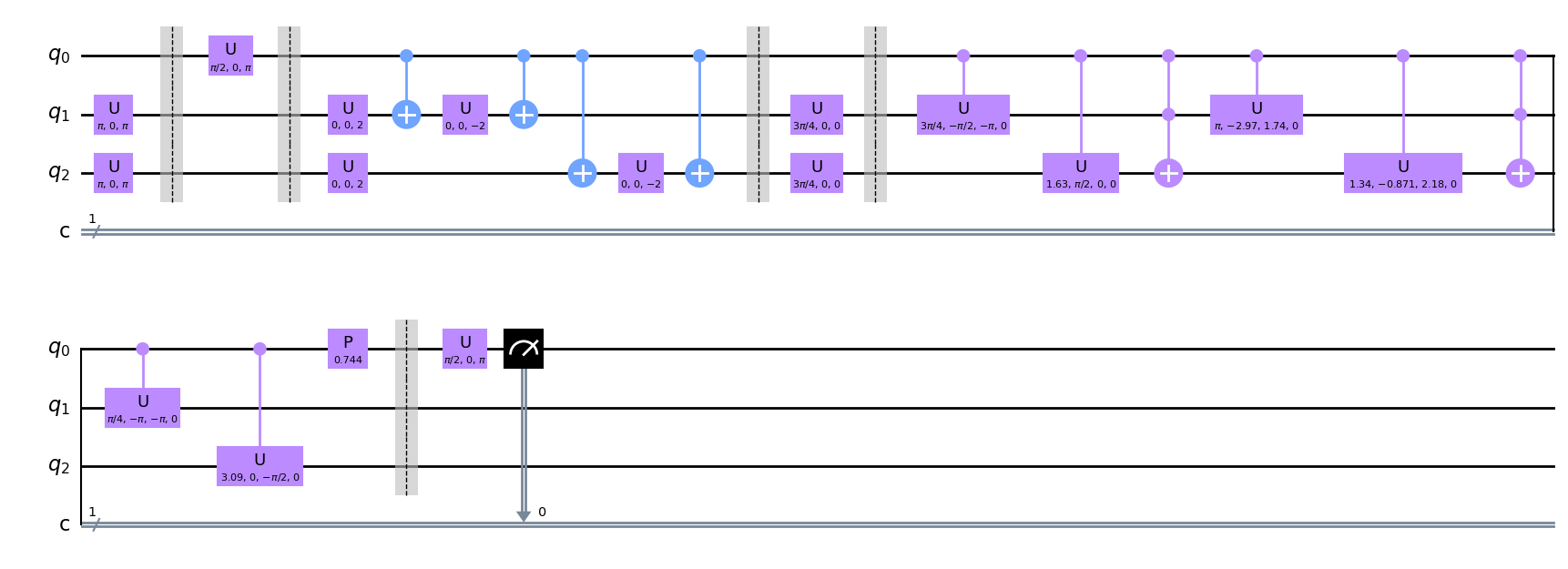}
  \caption{An example of the forward process circuits. After applying X gate on the bottom two qubits and Hadamard gate on the ancilla qubit, controlled $\exp(-i H_0)$ gate, $U$ gate, controlled $\exp(i G_{\pi/4})$ gate are implemented. Note that $P$ is a phase gate and $U$ is a custom unitary gate, which are defined by $P(\lambda) =\begin{pmatrix} 1 & 0 \\ 0 & e^{i\lambda}\end{pmatrix}$ and $U(\theta,\phi,\lambda) = \begin{pmatrix}
\cos(\frac{\theta}{2})&-e^{i \lambda}\sin(\frac{\theta}{2})\\
e^{i \phi}\sin(\frac{\theta}{2})&e^{i(\phi + \lambda)}\cos(\frac{\theta}{2})
\end{pmatrix}$ }
  \label{fig:supp:ForwardCircuit}
\end{figure}

To compute the characteristic function of the forward process $\widetilde{C}_f(1)$, we need to run $4\times 2 = 8$ quantum circuits because of the decomposition of ${\rho}_{0}^{\eq}$ into the four orthogonal states and the measurement of the ancilla qubit on two different bases. Figure.~\ref{fig:supp:ForwardCircuit} is the quantum circuit for computing the real part of $\tr{U\ad  e^{iG_{\tau}} U e^{-iH_0}\dya{E_4}}$.

\subsection{Backward process}
\begin{figure}[htp!]
  \centering
  \includegraphics[width=1 \textwidth]{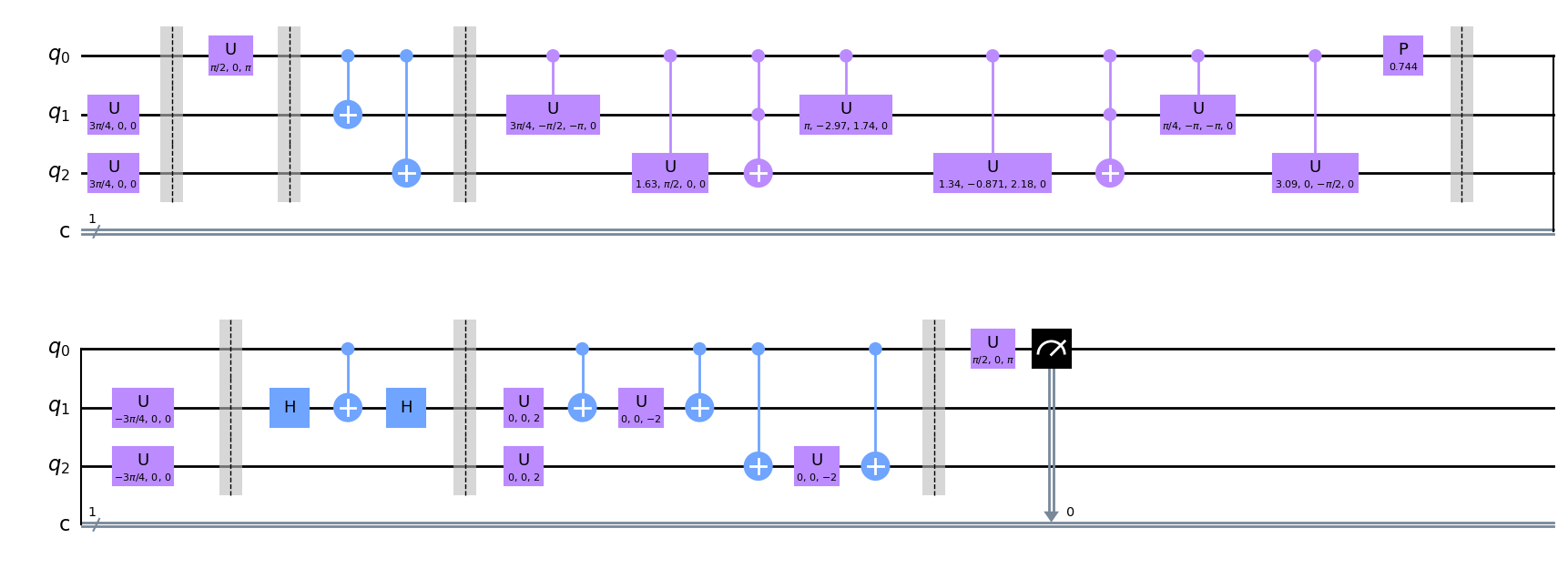}
  \caption{An example of the backward process circuits. After applying $U$ on the two qubits and Hadamard gate on the ancilla qubit, we add controlled $X \otimes X$ gate, controlled $\exp(i G_{\pi/4})$ gate, $U\ad$ gate, controlled $Z \otimes \id$ gate, and controlled $\exp(-i  H_0)$ gate. Note that H in the circuit is Hadamard gate.}
  \label{fig:supp:BackwardCircuit}
\end{figure}

Next, we explain the details of computing $\widetilde{C}_b(-1+0.5i)$. By computing $\alpha_k^{(0)}= \frac{1}{4} \tr{e^{-\beta H_0}\sigma_k}$ and $\alpha_k^{(\pi/4)}=\frac{1}{4} \tr{e^{0.5 G_{\pi/4}}\sigma_k}$, the decomposition of $\exp(0.5 H_0)$ and $\exp(-0.5 G_\tau)$ with the Pauli strings are given by 
\begin{equation}
\begin{split}
e^{-0.5 H_0} &= 2.3811( \id \otimes \id )
-1.81343( \id \otimes Z )
-1.81343( Z \otimes \id )
+1.3811( Z \otimes Z )\\
e^{0.5 G_\tau} &= 1.03141( \id \otimes \id) 
+ 0.126306( X \otimes X )
-0.126306( X \otimes Z )
-0.126306( Z \otimes X )
+0.126306( Z \otimes Z ).
\end{split}
\end{equation}
Therefore, we have $5\times 4=20$ non-zero values of $F_{k\ell}$. Because $\widetilde{\rho}_{\pi/4}$ is the linear combination of four orthogonal states, taking into account the fact that we need to perform the measurement on two different bases, we finally need to make $4 \times 5 \times 4 \times 2 = 160$ quantum circuits to compute $\widetilde{C}_b(-1+0.5i)$. Figure.~\ref{fig:supp:BackwardCircuit} is the quantum circuit for computing the imaginary part of $\tr{U e^{-iH_0}\sigma_k U^\dagger e^{iG_{\pi/4}}\sigma_{\ell} U\dya{E_1}U\ad }$, where $\sigma_k= Z \otimes \id$ and $\sigma_{\ell}=X \otimes X$.

\section{Detailed information about IBM machines}
\begin{figure}[htp!]
  \centering
  \includegraphics[width=0.25 \textwidth]{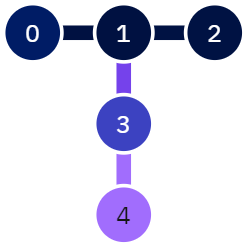}
  \caption{Qubit configuration of ibm\_lima.}
\label{fig:QbitConfigLima}
\end{figure}

\begin{figure}[htp!]
  \centering
  \includegraphics[width=0.25 \textwidth]{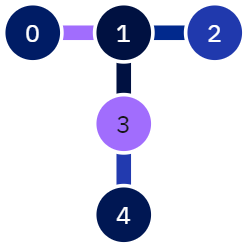}
  \caption{Qubit configuration of ibm\_belem.}
  \label{fig:QbitConfigBelem}
\end{figure}

\begin{figure}[htp!]
  \centering
  \includegraphics[width=0.25 \textwidth]{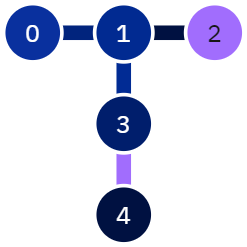}
  \caption{Qubit configuration of ibm\_quito.}
  \label{fig:QbitConfigQuito}
\end{figure}

\begin{figure}[htp!]
  \centering
  \includegraphics[width=0.25 \textwidth]{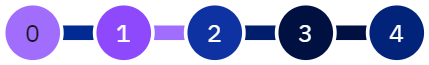}
  \caption{Qubit configuration of ibm\_manila.}
  \label{fig:QbitConfigManila}
\end{figure}

\begin{figure}[htp!]
  \centering
  \includegraphics[width=0.25 \textwidth]{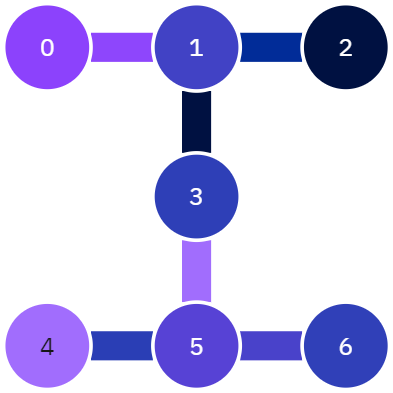}
  \caption{Qubit configuration of ibm\_lagos.}
  \label{fig:QbitConfigLagos}
\end{figure}

In this section, we present detailed information about the IBM quantum machines that we used for this {simulation}. 
We used ibmq\_lima, ibmq\_belem, ibmq\_quito, ibm\_oslo, ibmq\_manila, and ibm\_lagos. The parameters of these machines are shown in Tables.~\ref{table:ibmq_lima}, \ref{table:ibmq_belem}, \ref{table:ibmq_quito}, \ref{table:ibmq_manila}, and \ref{table:ibmq_lagos}. Note that ibm\_oslo has retired so that its information is no longer available. The qubit configurations of each machine are depicted in Figs.~\ref{fig:QbitConfigLima}, \ref{fig:QbitConfigBelem}, \ref{fig:QbitConfigQuito}, \ref{fig:QbitConfigManila} and \ref{fig:QbitConfigLagos}.

\begin{table}[htp!]
\begin{tabular}{|c||c|c|c|c|c|}

\hline
Qubit & 
$T_1$ ($\upmu$s) & 
$T_2$ ($\upmu$s) & 
Frequency (GHz) & 
Anharmonicity (GHz) &
Readout assignment error\\     

\hline
0 & 166.56 & 232.22 & 5.030 & -0.33574 & 2.130$\times 10^{-2}$  \\
1 & 140.20 & 135.21 & 5.128 & -0.31835 & 1.850$\times 10^{-2}$  \\
2 & 94.23  & 106.02 & 5.247 & -0.33360 & 1.860$\times 10^{-2}$  \\
3 & 86.82  & 95.33  & 5.303 & -0.33124 & 3.370$\times 10^{-2}$  \\
4 & 17.48  & 24.33  & 5.092 & -0.33447 & 4.780$\times 10^{-2}$ \\

\hline
Qubit & 
Prob meas 0  prep 1 & Prob meas 1 prep 0 & 
Readout length (ns) & 
ID error & 
$\sqrt{X}$ (SX) error \\

\hline
0 & 0.0338 & 0.0088 & 5912.889 & 2.946$\times 10^{-4}$ & 2.946$\times 10^{-4}$  \\
1 & 0.0256 & 0.0114 & 5912.889 & 4.617$\times 10^{-4}$ & 4.617$\times 10^{-4}$  \\
2 & 0.0288 & 0.0084 & 5912.889 & 5.618$\times 10^{-4}$ & 5.618$\times 10^{-4}$  \\
3 & 0.0442 & 0.0232 & 5912.889 & 2.428$\times 10^{-4}$ & 2.428$\times 10^{-4}$  \\
4 & 0.0760 & 0.0196 & 5912.889 & 8.183$\times 10^{-4}$ & 8.183$\times 10^{-4}$ \\
\hline

Qubit & 
Pauli-X error &  
CNOT error & Gate time (ns)\\

\cline{1-4}
0 & 2.946$\times 10^{-4}$ & 0\_1:0.00677 & 0\_1:305.778 \\
1 & 4.617$\times 10^{-4}$ & 1\_0:0.00677 & 1\_0:341.333 \\
  &           & 1\_3:0.01360 & 1\_3:497.778 \\
  &           & 1\_2:0.00677 & 1\_2:334.222 \\
2 & 5.618$\times 10^{-4}$ & 2\_1:0.00677 & 2\_1:298.667 \\
3 & 2.428$\times 10^{-4}$ & 3\_4:0.01679 & 3\_4:519.111 \\
  &           & 3\_1:0.01360 & 3\_1:462.222 \\
4 & 8.183$\times 10^{-4}$ & 4\_3:0.01679 & 4\_3:483.556 \\
\cline{1-4}

\end{tabular}
\caption{The parameter settings of ibmq\_lima. In this {simulation}, we used either Qubit 0, 1, 2, Qubit 0, 1, 3, Qubit 1, 2, 3, or Qubit 1, 3, 4.}
\label{table:ibmq_lima}
\end{table}
\begin{table}[htp!]
\begin{tabular}{|c||c|c|c|c|c|}

\hline
Qubit & 
$T_1$ ($\upmu$s) & 
$T_2$ ($\upmu$s) & 
Frequency (GHz) & 
Anharmonicity (GHz) &
Readout assignment error\\     

\hline
0 & 144.96 & 139.86 & 5.090 & -0.33612 & 1.990$\times 10^{-2}$  \\
1 & 85.46  & 94.85  & 5.246 & -0.31657 & 1.820$\times 10^{-2}$  \\
2 & 69.38  & 51.70  & 5.361 & -0.33063 & 2.510$\times 10^{-2}$  \\
3 & 102.99 & 117.70 & 5.170 & -0.33374 & 3.580$\times 10^{-2}$  \\
4 & 130.83 & 155.00 & 5.258 & -0.33135 & 1.900$\times 10^{-2}$ \\

\hline
Qubit & 
Prob meas 0  prep 1 & Prob meas 1 prep 0 & 
Readout length (ns) & 
ID error & 
$\sqrt{X}$ (SX) error \\

\hline
0 & 0.0330 & 0.0068 & 6158.222 & 1.737$\times 10^{-4}$ & 1.737$\times 10^{-4}$  \\
1 & 0.0320 & 0.0044 & 6158.222 & 2.221$\times 10^{-4}$ & 2.221$\times 10^{-4}$  \\
2 & 0.0436 & 0.0066 & 6158.222 & 2.661$\times 10^{-4}$ & 2.661$\times 10^{-4}$  \\
3 & 0.0564 & 0.0152 & 6158.222 & 3.742$\times 10^{-4}$ & 3.742$\times 10^{-4}$  \\
4 & 0.0292 & 0.0088 & 6158.222 & 5.762$\times 10^{-4}$ & 5.762$\times 10^{-4}$ \\
\hline

Qubit & 
Pauli-X error &  
CNOT error & Gate time (ns)\\

\cline{1-4}
0 & 1.737$\times 10^{-4}$ & 0\_1:0.01276 & 0\_1:810.667 \\
1 & 2.221$\times 10^{-4}$ & 1\_3:0.00776 & 1\_3:440.889 \\
  &           & 1\_2:0.00884 & 1\_2:419.556 \\
  &           & 1\_0:0.01276 & 1\_0:775.111 \\
2 & 2.661$\times 10^{-4}$ & 2\_1:0.00884 & 2\_1:384.000 \\
3 & 3.742$\times 10^{-4}$ & 3\_4:0.00978 & 3\_4:526.222 \\
  &           & 3\_1:0.00776 & 3\_1:405.333 \\
4 & 5.762$\times 10^{-4}$ & 4\_3:0.00978 & 4\_3:490.667 \\
\cline{1-4}

\end{tabular}
\caption{The parameter settings of ibmq\_belem. In this {simulation}, we used either Qubit 0, 1, 2, Qubit 0, 1, 3, Qubit 1, 2, 3, or Qubit 1, 3, 4.}
\label{table:ibmq_belem}
\end{table}

\begin{table}[htp!]
\begin{tabular}{|c||c|c|c|c|c|}

\hline
Qubit & 
$T_1$ ($\upmu$s) & 
$T_2$ ($\upmu$s) & 
Frequency (GHz) & 
Anharmonicity (GHz) &
Readout assignment error\\     

\hline
0 & 121.37 & 151.04 & 5.301 & -0.33148 & 5.870$\times 10^{-2}$ \\
1 & 43.25  & 79.69  & 5.081 & -0.31925 & 3.550$\times 10^{-2}$ \\
2 & 111.43 & 116.63 & 5.322 & -0.33232 & 6.710$\times 10^{-2}$ \\
3 & 86.75  & 17.79  & 5.164 & -0.33508 & 4.000$\times 10^{-2}$ \\
4 & 89.82  & 107.98 & 5.052 & -0.31926 & 3.290$\times 10^{-2}$ \\

\hline
Qubit & 
Prob meas 0  prep 1 & Prob meas 1 prep 0 & 
Readout length (ns) & 
ID error & 
$\sqrt{X}$ (SX) error \\

\hline
0 & 0.0838 & 0.0336 & 5351.111 & 4.604$\times 10^{-4}$ & 4.604$\times 10^{-4}$ \\
1 & 0.0402 & 0.0308 & 5351.111 & 2.924$\times 10^{-4}$ & 2.924$\times 10^{-4}$ \\
2 & 0.0646 & 0.0696 & 5351.111 & 2.564$\times 10^{-4}$ & 2.564$\times 10^{-4}$ \\
3 & 0.0616 & 0.0184 & 5351.111 & 1.178e-03 & 1.178e-03 \\
4 & 0.0466 & 0.0192 & 5351.111 & 3.567$\times 10^{-4}$ & 3.567$\times 10^{-4}$ \\
\hline

Qubit & 
Pauli-X error &  
CNOT error & Gate time (ns)\\

\cline{1-4}
0 & 4.604$\times 10^{-4}$ & 0\_1:0.01376  & 0\_1:234.667    \\
1 & 2.924$\times 10^{-4}$ & 1\_3:0.00920  & 1\_3:334.222    \\
  &           & 1\_2:0.00675  & 1\_2:298.667    \\
  &           & 1\_0:0.01376  & 1\_0:270.222    \\
2 & 2.564$\times 10^{-4}$ & 2\_1:0.00675  & 2\_1:263.111    \\
3 & 1.178e-03 & 3\_4:0.01483 & 3\_4:277.333    \\
  &           & 3\_1:0.00920  & 3\_1:369.778    \\
4 & 3.567$\times 10^{-4}$ & 4\_3:0.01483  & 4\_3:312.889 \\
\cline{1-4}

\end{tabular}
\caption{The parameter settings of ibmq\_quito. In this {simulation}, we used either Qubit 0, 1, 2, Qubit 0, 1, 3, Qubit 1, 2, 3, or Qubit 1, 3, 4.}
\label{table:ibmq_quito}
\end{table}

\begin{table}[htp!]
\begin{tabular}{|c||c|c|c|c|c|}

\hline
Qubit & 
$T_1$ ($\upmu$s) & 
$T_2$ ($\upmu$s) & 
Frequency (GHz) & 
Anharmonicity (GHz) &
Readout assignment error\\     

\hline
0 & 72.22  & 18.13 & 4.962 & -0.34463 & 4.840$\times 10^{-2}$  \\
1 & 176.31 & 63.37 & 4.838 & -0.34528 & 4.230$\times 10^{-2}$  \\
2 & 147.72 & 17.26 & 5.037 & -0.34255 & 2.810$\times 10^{-2}$  \\
3 & 161.33 & 60.76 & 4.951 & -0.34358 & 1.890$\times 10^{-2}$  \\
4 & 47.92  & 37.82 & 5.065 & -0.34211 & 2.360$\times 10^{-2}$  \\

\hline
Qubit & 
Prob meas 0  prep 1 & Prob meas 1 prep 0 & 
Readout length (ns) & 
ID error & 
$\sqrt{X}$ (SX) error \\

\hline
0 & 0.0778 & 0.0190 & 5351.111 & 5.700$\times 10^{-4}$ & 5.700$\times 10^{-4}$  \\
1 & 0.0456 & 0.0390 & 5351.111 & 3.243$\times 10^{-4}$ & 3.243$\times 10^{-4}$  \\
2 & 0.0380 & 0.0182 & 5351.111 & 2.435$\times 10^{-4}$ & 2.435$\times 10^{-4}$  \\
3 & 0.0250 & 0.0128 & 5351.111 & 1.617$\times 10^{-4}$ & 1.617$\times 10^{-4}$  \\
4 & 0.0352 & 0.0120 & 5351.111 & 4.111$\times 10^{-4}$ & 4.111$\times 10^{-4}$  \\
\hline

Qubit & 
Pauli-X error &  
CNOT error & Gate time (ns)\\

\cline{1-4}
0 & 5.700$\times 10^{-4}$ & 0\_1:0.00800 & 0\_1:277.333 \\
1 & 3.243$\times 10^{-4}$ & 1\_2:0.01801 & 1\_2:469.333 \\
  &           & 1\_0:0.00800 & 1\_0:312.889 \\
2 & 2.435$\times 10^{-4}$ & 2\_3:0.00652 & 2\_3:355.556 \\
  &           & 2\_1:0.01801 & 2\_1:504.889 \\
3 & 1.617$\times 10^{-4}$ & 3\_4:0.00499 & 3\_4:334.222 \\
  &           & 3\_2:0.00652 & 3\_2:391.111 \\
4 & 4.111$\times 10^{-4}$ & 4\_3:0.00499 & 4\_3:298.667 \\
\cline{1-4}

\end{tabular}
\caption{The parameter settings of ibmq\_manila. In the {simulation}, we used wither Qubit 0, 1, 2, Qubit 1, 2, 3, or Qubit 2, 3, 4.}
\label{table:ibmq_manila}
\end{table}

\begin{table}[htp!]
\begin{tabular}{|c||c|c|c|c|c|}

\hline
Qubit & 
$T_1$ ($\upmu$s) & 
$T_2$ ($\upmu$s) & 
Frequency (GHz) & 
Anharmonicity (GHz) &
Readout assignment error\\     

\hline
0 & 120.31 & 43.15  & 5.235 & -0.33987 & 1.860$\times 10^{-2}$ \\
1 & 134.21 & 86.21  & 5.100 & -0.34325 & 1.520$\times 10^{-2}$ \\
2 & 198.82 & 137.93 & 5.188 & -0.34193 & 6.700e-03 \\
3 & 253.10 & 89.53  & 4.987 & -0.34529 & 1.400$\times 10^{-2}$ \\
4 & 66.93  & 35.95  & 5.285 & -0.33923 & 2.260$\times 10^{-2}$ \\
5 & 126.12 & 65.55  & 5.176 & -0.34079 & 1.620$\times 10^{-2}$ \\
6 & 189.10 & 107.66 & 5.064 & -0.34276 & 1.420$\times 10^{-2}$ \\

\hline
Qubit & 
Prob meas 0  prep 1 & Prob meas 1 prep 0 & 
Readout length (ns) & 
ID error & 
$\sqrt{X}$ (SX) error \\

\hline
0 & 0.0138 & 0.0234 & 789.333 & 2.333$\times 10^{-4}$ & 2.333$\times 10^{-4}$ \\
1 & 0.0174 & 0.0130 & 789.333 & 1.712$\times 10^{-4}$ & 1.712$\times 10^{-4}$ \\
2 & 0.0056 & 0.0078 & 789.333 & 2.233$\times 10^{-4}$ & 2.233$\times 10^{-4}$ \\
3 & 0.0166 & 0.0114 & 789.333 & 1.805$\times 10^{-4}$ & 1.805$\times 10^{-4}$ \\
4 & 0.0208 & 0.0244 & 789.333 & 2.025$\times 10^{-4}$ & 2.025$\times 10^{-4}$ \\
5 & 0.0180 & 0.0144 & 789.333 & 1.640$\times 10^{-4}$ & 1.640$\times 10^{-4}$ \\
6 & 0.0146 & 0.0138 & 789.333 & 2.433$\times 10^{-4}$ & 2.433$\times 10^{-4}$ \\
\hline

Qubit & 
Pauli-X error &  
CNOT error & Gate time (ns)\\

\cline{1-4}
0 & 2.333$\times 10^{-4}$ & 0\_1:0.00783 & 0\_1:576.000 \\
1 & 1.712$\times 10^{-4}$ & 1\_3:0.00453 & 1\_3:334.222 \\
  &           & 1\_2:0.00559 & 1\_2:327.111 \\
  &           & 1\_0:0.00783 & 1\_0:611.556 \\
2 & 2.233$\times 10^{-4}$ & 2\_1:0.00559 & 2\_1:291.556 \\
3 & 1.805$\times 10^{-4}$ & 3\_1:0.00453 & 3\_1:298.667 \\
  &           & 3\_5:0.00879 & 3\_5:334.222 \\
4 & 2.025$\times 10^{-4}$ & 4\_5:0.00642 & 4\_5:362.667 \\
5 & 1.640$\times 10^{-4}$ & 5\_4:0.00642 & 5\_4:327.111 \\
  &           & 5\_6:0.00690 & 5\_6:256.000 \\
  &           & 5\_3:0.00879 & 5\_3:298.667 \\
6 & 2.433$\times 10^{-4}$ & 6\_5:0.00690 & 6\_5:291.556 \\
\cline{1-4}

\end{tabular}
\caption{The parameter settings of ibmq\_lagos. In the {simulation}, we used either Qubit 0, 1, 2, Qubit 0, 1, 3, Qubit 1, 2, 3, Qubit 1, 3, 5, Qubit 3, 4, 5, or Qubit 3, 5, 6.}
\label{table:ibmq_lagos}
\end{table}

\end{document}